# Breaking symmetry with light: photo-induced chirality in a non-chiral crystal

Z. Zeng[1,2], M. Först[1], M. Fechner[1], M. Buzzi[1], E. Amuah[1], C. Putzke[1], P.J.W. Moll[1], D. Prabhakaran[2], P. Radaelli[2], A. Cavalleri[1,2]

[1]Max Planck Institute for the Structure and Dynamics of Matter, Hamburg, Germany
[2]Department of Physics, Clarendon Laboratory, University of Oxford, Oxford, United Kingdom

**Chirality is a pervasive form of symmetry that is intimately connected to the physical properties of solids, as well as the chemical and biological activity of molecular systems. However, its control with light is challenging, because inducing chirality in a non-chiral material requires that all mirrors and all roto-inversions be simultaneously broken. Electromagnetic fields exert only oscillatory forces that vanish on average, mostly leading to entropy increase that does not break symmetries, per se. Here, we show that chirality of either handedness can be generated in the non-chiral piezoelectric material $BPO_4$, in which two compensated sub-structures of opposite handedness coexist within the same unit cell. By resonantly driving either one of two orthogonal, doubly degenerate vibrational modes at Terahertz frequency, we rectify the lattice distortion and exert a displacive force onto the crystal. The staggered chirality is in this way uncompensated in either direction, inducing chiral structure with either handedness. The rotary power of the photo-induced phases is comparable to the static value of prototypical chiral $\alpha$-quartz, limited by the strength of the pump laser pulse.**

An object is defined as chiral if its mirror image cannot be superimposed to itself through any combination of rotations or translations. In crystalline systems, the structural chirality is predetermined by the lattice structure during the formation process(*1*), making it challenging to manipulate the handedness of the system after growth. For example, the chiral crystal $\alpha$-quartz(*2*) (space group $P3_121$ and $P3_221$) can exist in either left- or right-handed structures, characterized by atomic spirals of opposite handedness within the unit cell (Fig. 1a). Once formed, chiral crystals of opposite handedness cannot be switched into each other without melting and recrystallization of the material(*3, 4*).

Antiferro-chirals are a distinct class of achiral systems, in which the unit cell comprises chiral fragments with opposite handedness(*5*). The overall system remains achiral due to the degeneracy between the left- and right-handed structures, resembling racemic crystals(*6*). One notable feature of these systems is their potential to develop chirality under external perturbations that can uncompensate the staggered chiral fragments.

Boron phosphate ($BPO_4$, space group I-4) is as a representative example of antiferro-chiral materials. Its equilibrium lattice and the left- and right-handed chiral sub-structures are sketched in Figure 1b(*7*). The displacement of the atomic structure along the coordinates of B-symmetry modes (with amplitudes $Q_B$) lifts the degeneracy between the local structures of opposite handedness, resulting in a ferri-chiral state. Figure 2a highlights this behavior, where the shaded atomic motions along the B-mode coordinates enhance and reduce the amplitudes of the left- and right-handed local chiral structures, respectively. As a result, the handedness of the ferri-chiral state can be controlled by the direction of the phonon displacement(*8*), i.e. the sign of its amplitude $Q_B$, providing an opportunity to rationally design the chirality in these systems by structural engineering.

This effect is not easily stimulated with external fields, as one needs to couple to a specific lattice mode which needs to be displaced in a specific direction. This is achieved transiently through nonlinear phononics(*9-17*), an effective approach to coherently control the atomic structure with light. This concept provides a novel basis for the rational design of crystal structures and symmetries with light, inducing desirable functional properties at high speed. In one specific type of nonlinear phononic interaction, the square of a selectively driven, infrared-active THz-frequency phonon mode $Q_{IR}$ couples linearly to a second mode $Q_2$, inducing a rectified, displacive force along the normal mode coordinate $Q_2$. This rectified force induces a transient crystal structure not accessible at equilibrium(*18-22*).

In BPO$_4$, a displacive force on the B-symmetry phonon modes that control chirality can be achieved by driving either of the doubly degenerate infrared-active E-symmetry phonon modes, polarized along the *a* and *b* axes respectively. According to the lowest-order coupling term of the form $U = -\alpha Q_{E,a}^2 Q_B + \alpha Q_{E,b}^2 Q_B$ (see Supplementary Information), the coherent drive of phonon mode $Q_{E,a}$ by a resonant THz frequency field exerts a rectified force onto $Q_B$ in the positive direction, leading to a positive transient displacement of the lattice along the B-mode coordinates away from equilibrium (Fig. 2b). Conversely, if the orthogonal mode $Q_{E,b}$ is resonantly driven, the transient displacement along $Q_B$ changes direction due to the opposite sign in the coupling term (Fig. 2c). Hence, the system can be driven into either one of the two opposite chiral states by controlling the polarization of the THz frequency excitation pulse.

This effect can be simulated for BPO$_4$ through two coupled equations of motion for the resonantly driven doubly-degenerate mode $Q_{E,a/b}$ and the set of four anharmonically coupled B-symmetry modes $Q_{B,i}$ ($i = 1 \dots 4$), taking the form

$$(1) \quad \frac{\partial^2}{\partial t^2} Q_{E,a/b}(t) + 2\gamma_{E,a/b} \frac{\partial}{\partial t} Q_{E,a/b}(t) + \omega_{E,a/b}^2 Q_{E,a/b}(t) = Z_{E,a/b}^* E(t)$$

$$(2) \quad \frac{\partial^2}{\partial t^2} Q_{B,i}(t) + 2\gamma_{B,i} \frac{\partial}{\partial t} Q_{B,i}(t) + \omega_{B,i}^2 Q_{B,i}(t) = +/- \alpha Q_{E,a/b}^2$$

Here, $\gamma_{E,a/b}$ and $\gamma_{B,i}$ are the damping coefficients and $\omega_{E,a/b}$ and $\omega_{B,i}$ the frequencies of the phonon modes. $Z_{E,a/b}^*$ is the effective charge that couples the infrared-active $Q_{E,a/b}$ modes to the pulsed THz electric field $E(t) = E_0 \sin(\omega_{B,i} t) e^{-\frac{t^2}{2\tau^2}}$. Importantly, the sign of the force on the B-symmetry modes (right-hand side in Eq. (2)) depends on whether the E-symmetry mode is excited along the *a* or the *b* axis. We used ab-initio calculations to determine all the relevant phonon parameters used in these equations (see Supplementary Information). These calculations predict that transient displacement of the crystal along the four B-symmetry phonons drives the system into a chiral state, which is determined by the linear superposition of these modes (see Supplementary Information).

The displacement causes two optical effects on a time delayed probe pulse, that is optical activity, which relates to the wanted induced chirality, and induced birefringence, which can be subtracted from the total optical signal to reveal the optical activity. Figures 3a and 3b show the time dependent changes of each one of these properties (optical activity and induced birefringence) for excitation of the mode $Q_{E,a}$ along the *a* axis, as calculated from the transient lattice structure and taking into account the B-mode dependent changes in the diagonal and off-diagonal elements of the optical permittivity. The latter were determined by an ab-initio density functional theory approach (see Supplementary Information).

The optical activity alone generates a characteristic polarization rotation which is *independent* on the incident polarization of the probe light in the *a-b* plane(*23, 24*) (see inset of Fig. 3a). The birefringence, on the other hand, introduces a modulation of the polarization rotation response with a four-fold symmetry with respect to the incident probe polarization (see inset of Fig. 3b). The total polarization rotation signal follows the form

$$\theta(\varphi) = A_1 \rho + A_2 \sin(4\varphi - \phi)(\Delta n)^2 , \quad (3)$$

where $\varphi$ is the relative angle between the pump and the probe polarization, $\phi$ is the angle between the pump polarization and the optical axis of the transient birefringence, $\rho$ is the rotary power proportional to the optical activity, and $\Delta n$ is the birefringence-induced difference in refractive index. The overall time dependent polarization rotation signal, calculated as a function of the incident polarization, is shown in Fig. 3c for a THz peak electric field of 5 MV/cm.

When the polarization of the pump is oriented along the $b$ axis to excite the mode $Q_{E,b}$, the induced displacement of the B-symmetry modes changes direction. Hence, both the optical activity and the induced birefringence are reversed (Fig. 3d & 3e), resulting in a sign change of the overall polarization rotation signal compared to the excitation along the $a$-axis (Fig. 3f).

Experimental validation of these predictions was obtained using the optical setup sketched in Figure 4a. The BPO$_4$ sample, held at room temperature, was excited by 19-THz center frequency pulses of 3 THz full width at half maximum, with a maximum excitation fluence of 5.0 mJ/cm$^2$ and a corresponding peak electric field of 5.1 MV/cm. These pulses were linearly polarized along either the $a$ or the $b$ axis, resonantly driving each of the doubly degenerate E-symmetry phonon mode at its 18.9 THz transverse-optical frequency (see Methods and Supplementary Information).

As shown in Figure 4b, a time-dependent rotation of the probe polarization was induced by the phonon excitation with a pump polarized along the crystal $a$ axis. At each probe polarization angle, we found a sudden onset of a rotation around time zero, followed by a decay lasting few-picoseconds, far longer than the 200 femtoseconds duration of the excitation pulse. The signal displayed a 90-degree periodicity with the incident probe polarization due to the transient birefringence discussed above.

Since the modulation induced by the birefringence averages to zero over all the probe polarizations(*25, 26*), we can extract the transient optical activity by averaging the signal over all the incident probe polarizations at each time delay (Fig. 4c). The result shows a finite and positive signal, providing clear evidence for a non-equilibrium chiral state. Its lifetime follows the excitation and decay of the resonantly driven optical phonon.

We further rotated the pump polarization by 90° to resonantly drive the E-symmetry phonon along the BPO$_4$ crystal $b$-axis. The corresponding time-dependent polarization rotation, again as a function of the incident probe polarization, and the corresponding optical activity are shown in Figs. 4d and 4e. As predicted, the signals reversed sign, evidencing opposite handedness of the light-induced ferri-chiral state compared to the $a$-axis excitation.

Further characterization of the light-induced chiral state validates the predicted nonlinear phononic mechanism. Figure 4a shows the rotary power as a function of the peak electric fields of the 19-THz excitation pulse, evidencing a quadratic field dependence and a sign reversal for the two different pump polarizations. This is consistent with the nonlinear phonon interaction potential $U = -\alpha Q_{E,a}^2 Q_B + \alpha Q_{E,b}^2 Q_B$. In addition, the magnitude of the non-equilibrium rotary power was resonantly enhanced when the excitation pulses were tuned to the 18.9 THz transverse optical frequency of the doubly degenerate E-symmetry phonon (Fig. 4b). We estimate the magnitude of the light-induced optical activity in BPO$_4$ at resonance by comparing the non-equilibrium rotary power to the static value of $\alpha$-quartz, a commonly used material for polarization rotation in optics(*27*). For the pump fluence available in the experiment, and assuming that the chiral state is induced only within the extinction depth $\delta$ of the excitation pulses (see Supplementary Information), the light-induced rotary power of BPO$_4$ is comparable to the equilibrium value of $\alpha$-quartz.

Extension of this approach to ferri-chiral systems may enable ultrafast switching with the nonlinear phononic protocol discussed here. Applications to ultrafast memory devices would follow, as well as to more sophisticated optoelectronic platforms, powered by light and connected to the handedness of matter. More broadly, the emergence of chirality on the ultrafast time scale, together with the ability to switch between chirality of opposite handedness, offers exciting opportunities for exploring new phenomena in out-of-equilibrium physics of complex matter, especially in topological(*28-30*) and correlated systems(*31, 32*) where handedness plays an important role.

**Acknowledgements:**

This work received funding from the Cluster of Excellence 'CUI: Advanced Imaging of Matter' of the Deutsche Forschungsgemeinschaft (DFG), EXC 2056, project ID 390715994. We thank X. Deng and N. Taherian for help with the sample preparation and the optical setup, respectively, X. Wang for his support in the implementation of the optical activity calculations, P. Licht for technical assistance, and J. Harms for assistance with graphics.

**Figures:**

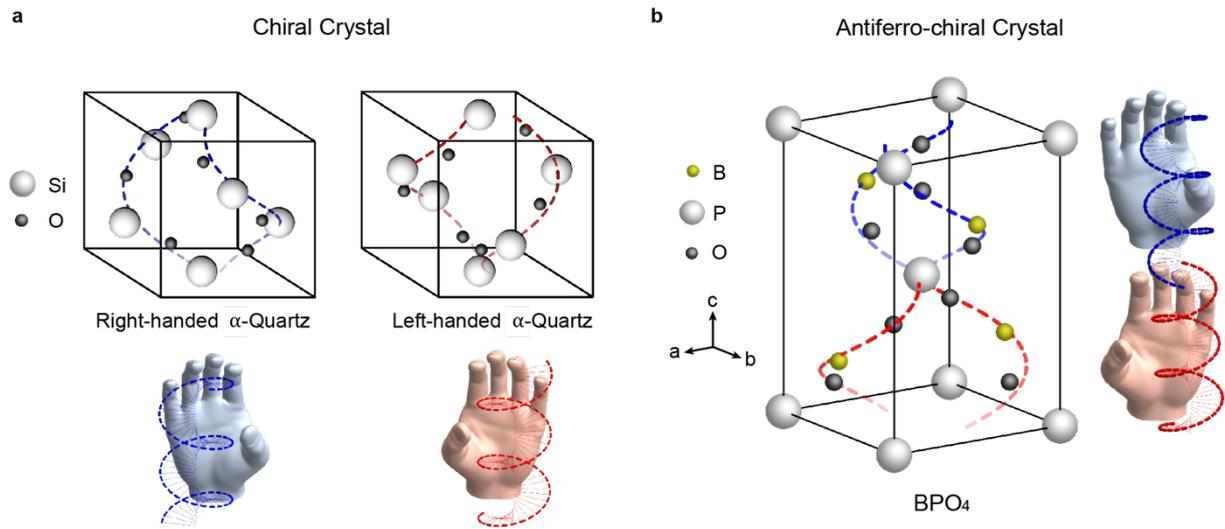

*Figure 1. | **Chirality in the solid state.** (a) The prototypical chiral crystal α-quartz exists in left- and right-handed configuration, determined by the spiral atomic structure formed during the growth process. (b) The unit cell of the antiferro-chiral crystal BPO₄ is composed of chiral sub-structures of opposite handedness. The degeneracy of these left- and right-handed structures makes the overall system achiral.*

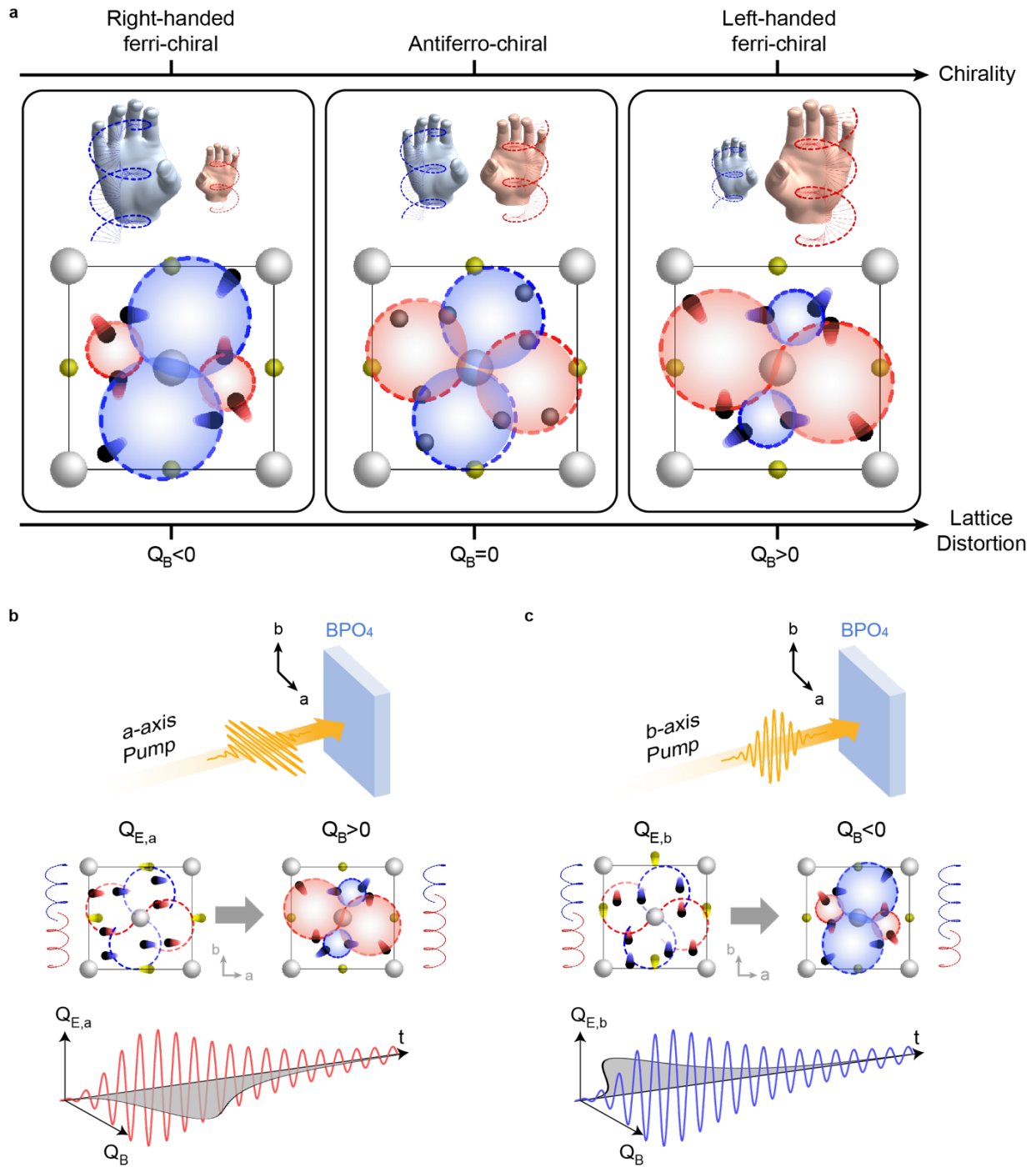

*Figure 2. | Light-induced chirality in antiferro-chiral BPO₄.* *(a) The atomic displacements in BPO$_4$ along B-symmetry phonons lifts the degeneracy between the local structures of left and right handedness, driving the system from the antiferro-chiral into a ferri-chiral state. The phonon displacement in the opposite direction induces chirality of opposite handedness. (b) A THz pump electric field polarization along the a axis induces coherent oscillations of the mode $Q_{E,a}$ about its equilibrium position. A positive transient displacement along the B mode coordinates is induced via nonlinear phonon coupling, driving the system into the non-equilibrium ferri-chiral state with left handedness. (c) Exciting the doubly generate E-symmetry phonon along the b axis induces coherent oscillations of the $Q_{E,b}$ mode about its equilibrium position. A negative transient displacement along the $Q_B$ mode coordinates is induced via nonlinear phonon coupling, driving the system into ferri-chiral state with right handedness.*

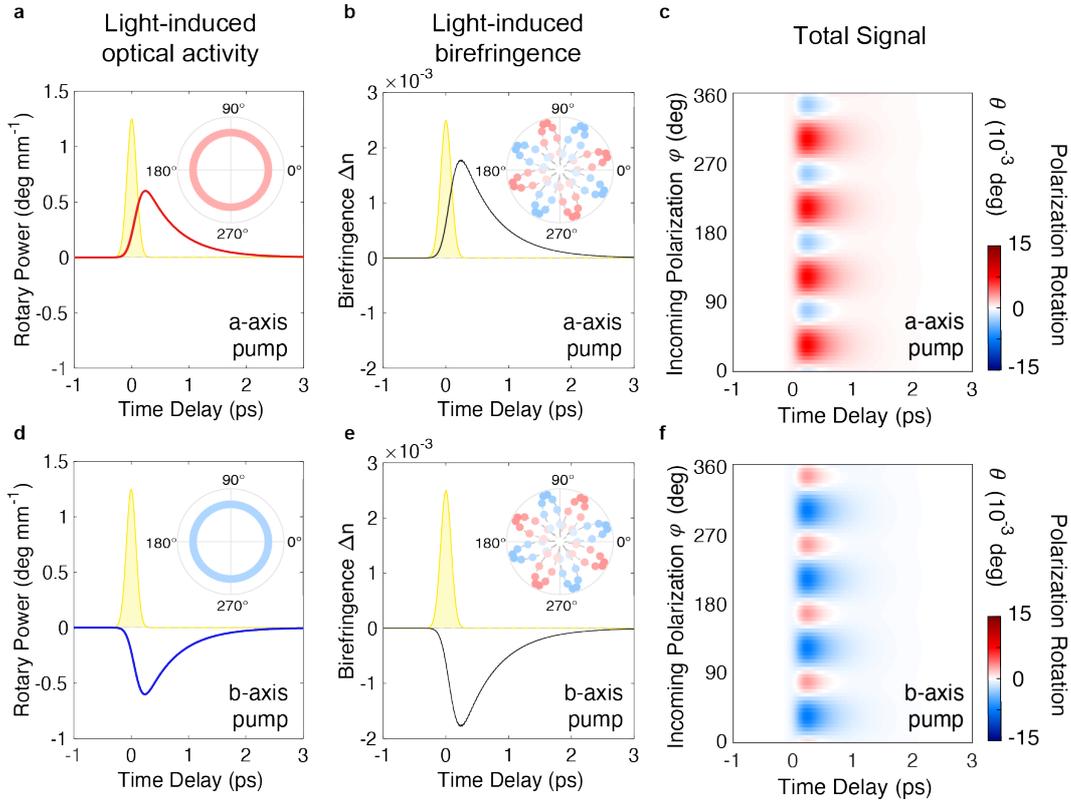

*Figure 3. | Theoretical calculations of light-induced chirality.* (a) Light-induced optical activity for the a-axis pump as a function of pump-probe time delay. The yellow shaded area is the temporal profile of the excitation pulse. Inset: the corresponding amplitude of the polarization rotation signal as a function of incident polarization. (b) Light-induced birefringence for the same a-axis pump as a function of pump-probe time delay. Inset: the corresponding amplitude modulation of the polarization rotation signal as a function of the incident polarization. (c) The resulting total polarization rotation signal for the a-axis pump as a function of incident polarization and pump-probe time delay. (d) Same as panel (a) for excitation along the b axis. (e) Same as panel (b) for excitation along the b axis (f) Same as panel (c) for excitation along the b axis. For all these calculations, we used the 19-THz pump pulse of 5 MV/cm peak electric field.

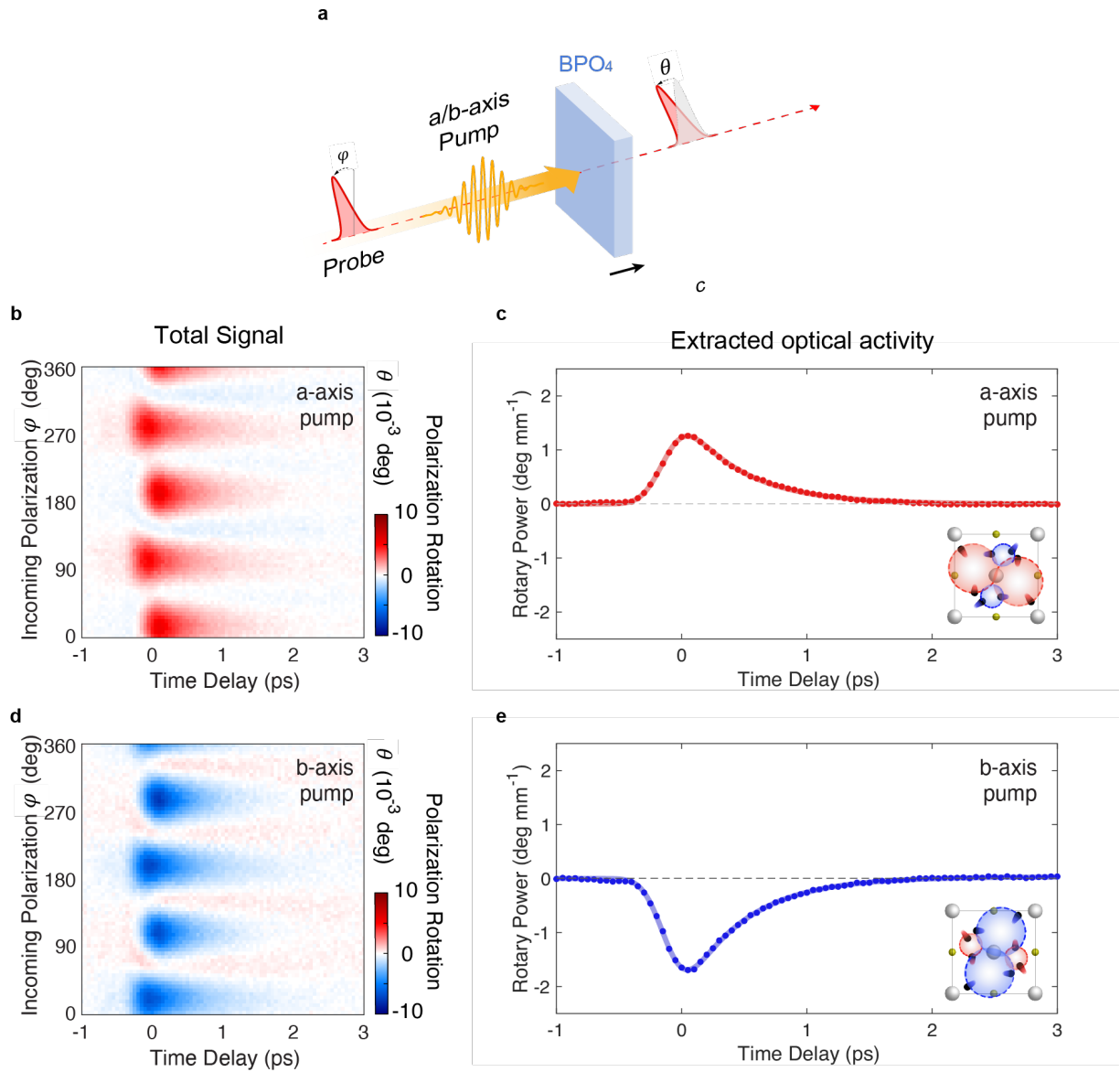

**Figure 4. | Time-resolved polarization rotation measurements.** *(a) Schematic of the pump-probe experiment. A linearly polarized mid-infrared pulse, polarized along either the a- or b-axis, is used to drive the BPO$_4$ crystal into chiral states. A time delayed near-infrared pulse probes the state by the measurement of its polarization rotation, carried out as a function of the probe incident polarization. (b) Time delay dependent polarization rotation signal for a-axis excitation as a function of probe incident polarization. (c) Time delay dependent rotary power, proportional to the optical activity, extracted from the data shown in panel (b) and considering the finite extinction depth δ of the excitation pulses. (d) Same as panel (b) for b-axis excitation. (e) Same as panel (c) for b-axis excitation.*

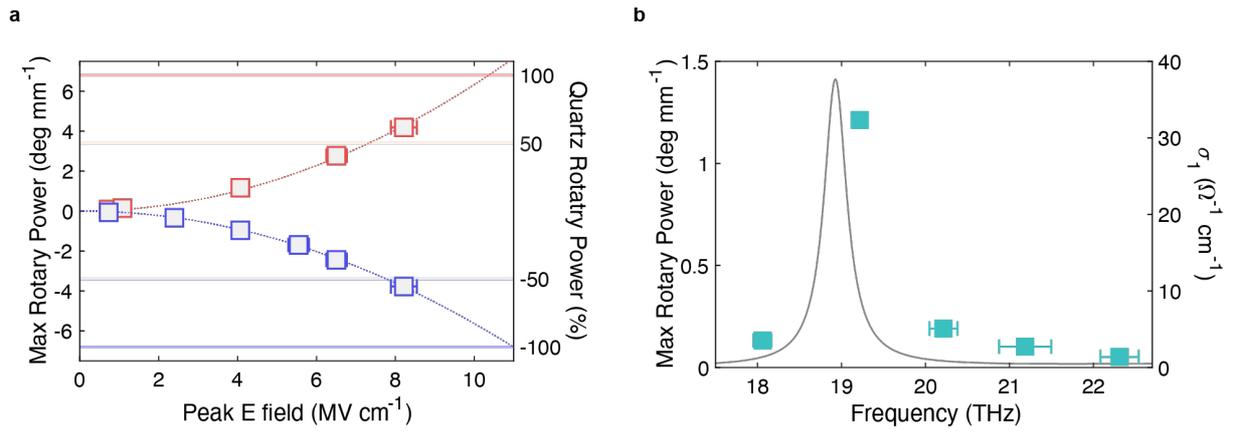

***Figure 5. | Characterization of the light-induced ferri-chiral states.*** *(a) The rotary power of the transient state as a function of the THz pulse peak electric field for a-axis polarized (red) and b-axis polarized (blue) excitation. (b) Maximum of rotary power at fixed peak electric field of 5.1 MV/cm as a function of the center frequency of the excitation pulse. The horizontal error bars are the 1σ confidence interval of the pump center frequency. Gray curve, real part of the optical conductivity.*

# Supplementary Materials for
# Breaking symmetry with light: photo-induced chirality in a non-chiral crystal

Z. Zeng[1,2], M. Först[1], M. Fechner[1], M. Buzzi[1], E. Amuah[1], C. Putzke[1], P.J.W. Moll[1], D. Prabhakaran[2], P. Radaelli[2], A. Cavalleri[1,2]

[1]Max Planck Institute for the Structure and Dynamics of Matter, Hamburg, Germany
[2]Department of Physics, Clarendon Laboratory, University of Oxford, Oxford, United Kingdom

**S1. Experimental Setup**

The schematic of the pump-probe setup used in the experiment is shown in Figure S1. The THz excitation pulses were generated by difference frequency generation (DFG) in a GaSe crystal, using the independently wavelength tunable near-infrared signal outputs of two optical parametric amplifiers (OPAs). The OPAs were seeded by the same white light and pumped by overall 1.5-mJ, 30-fs pulses at 800 nm wavelength from a 1-kHz repetition rate Ti:sapphire amplifier. An off-axis parabolic mirror was used to focus the THz beam onto the sample with a ~70 μm FWHM spot size. The generated THz pulses were carrier-envelope-phase (CEP) stable and characterized by electro-optical sampling in a second GaSe crystal at the sample position(*33*).

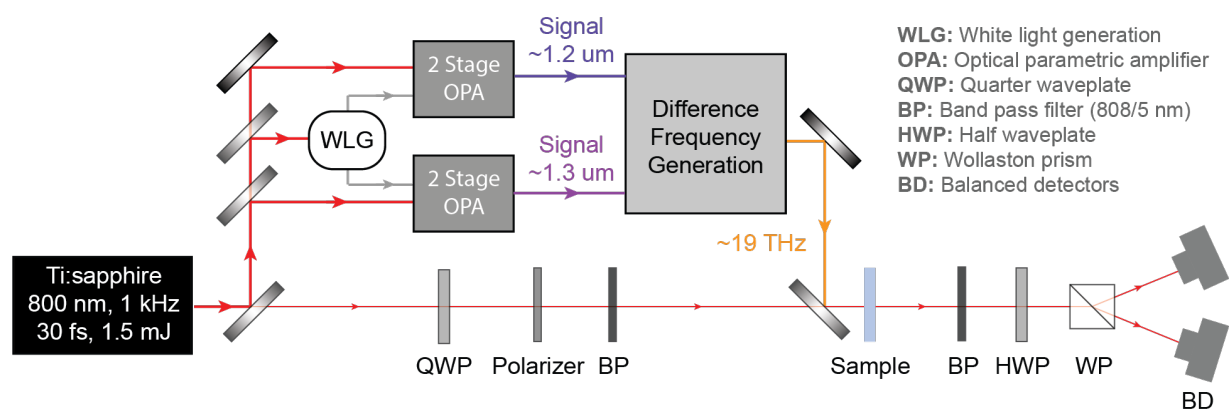

**Figure S2. | Schematic drawing of the experimental setup.**

The probe pulses for the polarization rotation measurements in the BPO$_4$ (001) single crystal were sourced by the same amplifier system and focused onto the sample with a lens to a ~30 μm spot. Polarization control of the probe incident on the sample was achieved using a quarter waveplate and a polarizer. Its pump-induced polarization rotation was measured using a combination of a half waveplate, a Wollaston prism and two photodiodes.

The pump and probe beams were aligned collinear and at normal incidence to the sample surface. Two identical narrow bandpass filters (center wavelength at 808 nm and bandwidth of 5 nm) were employed before and after the sample to filter out high-frequency components in the time-resolved response induced by the Pockels effect, which, in contrast to the light-induced birefringence and the light-induced optical activity, generates sidebands to the incident probe spectrum. The THz excitation pulses were linearly polarized along the *a* or *b* axis of the *c*-axis oriented $BPO_4$ crystal. All the measurements were carried out at room temperature.

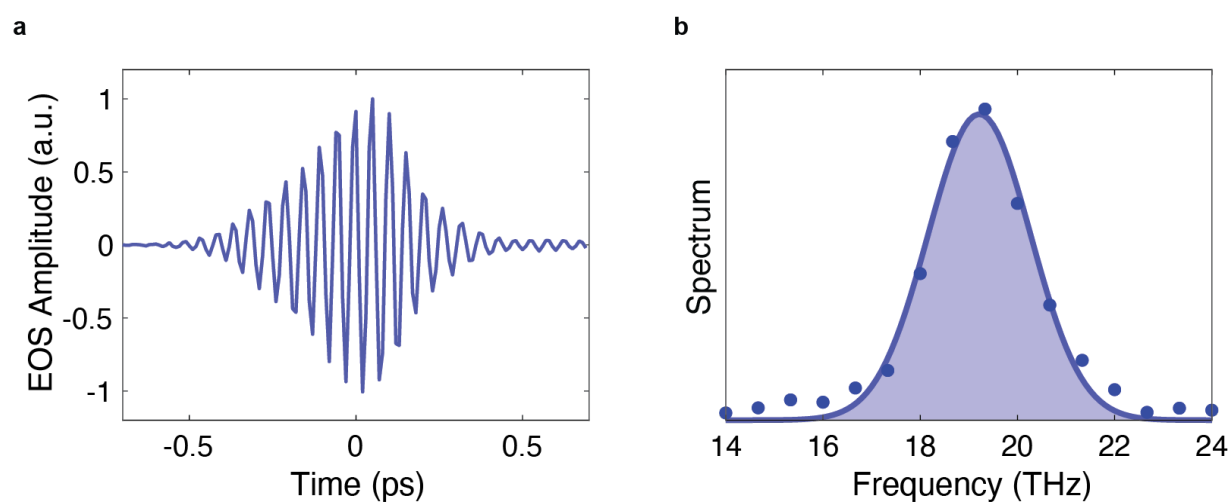

**Figure S2. | Characterization of the 19-THz pump pulse.** (a) Typical time-dependent electric field of the excitation pulses characterized by electro-optical sampling (EOS). (b) Corresponding frequency spectrum obtained by Fourier transformation.

## S2. Sample Preparation and Characterization

Polycrystalline $BPO_4$ powder was synthesized with high-purity $H_3BO_3$ (99.9995%) and $NH_4H_2PO_4$ (99.998%) powders, before being ground, pelletized, and sintered. Single crystals were grown from this powder using $Li_2MoO_4$ flux via the top-seeded solution growth technique in a platinum crucible at up to 975°C. After homogenization, the melt was slowly cooled down, resulting in the growth of high-quality $BPO_4$ single crystals.

These crystals were characterized by x-ray diffractometry and oriented with an optically flat (001) surface using focused ion beam milling. Xenon ions were accelerated at 30 kV at a beam current of 2.5 μA for the surface orientation. To reduce roughness, the sample was finally milled at grazing incidence with a beam current of 200 nA(*34*). The thickness of the sample used in this experiment was ~200 μm.

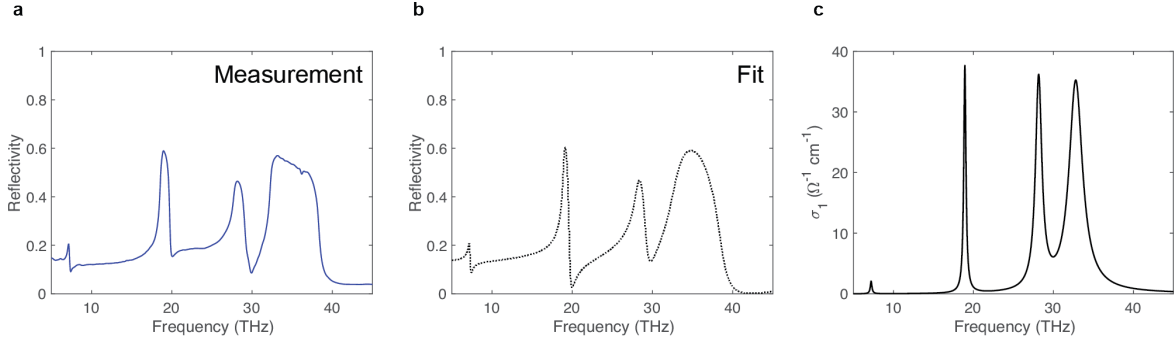

**Figure S3. | Characterization of the BPO$_4$ sample. (a) The static reflectivity spectrum of BPO$_4$ for electric field polarization along the *a*-axis. (b) Fit of the measured spectrum using the dielectric function for phonons. (c) Frequency dependent real part of the optical conductivity extracted from this fit.**

The static optical properties in the THz frequency range were characterized by FTIR reflectivity measurements, with the light electric field polarized along the *a* axis. Figure S3 clearly shows all the four E-symmetry phonon modes of BPO$_4$ in the detected spectrum, which was fitted with the following dielectric function for phonons(*35*).

$$\varepsilon(\omega) = \varepsilon_\infty + \varepsilon_\infty \sum_j \frac{\omega_{LO,j}^2 - \omega_{TO,j}^2}{\omega_{TO,j}^2 - \omega^2 - i\omega\gamma_j}$$

The fitting results are listed in Table S1. In the experiments reported here, we resonantly excited the E(2) mode.

| Phonon Mode | $\omega_{TO,j}$ (THz) | $\omega_{LO,j}$ (THz) | $\gamma_j$ (THz) | $\varepsilon_\infty$ |
|---|---|---|---|---|
| E(1) | 7.2 | 7.4 | 0.24 | / |
| E(2) | 18.9 | 20.3 | 0.33 | / |
| E(3) | 28.2 | 30.5 | 0.96 | / |
| E(4) | 32.8 | 37.0 | 2.0 | / |
| / | / | / | / | 2.70 |

**Table S1. | Fitting result of the E-symmetry phonons in BPO$_4$ polarized along the *a* and *b* axes.**

## S3. Analysis of the Time-resolved Polarization Rotation Measurements

In the time-resolved experiments, the probe pulses propagate along the optical axis of the crystal (c-axis), polarized in the *a-b* plane. Hence, the permittivity tensor elements along crystal *a* and *b* axis are considered as follows.

Due to the $\bar{4}$ symmetry of BPO$_4$, the static permittivity is isotropic in the *a-b* plane:

$$\boldsymbol{\varepsilon} = \begin{pmatrix} \varepsilon_{11} & 0 \\ 0 & \varepsilon_{11} \end{pmatrix}.$$

Based on a symmetry analysis, and as confirmed by our ab-initio calculations, the atomic displacements along the B-symmetry phonon modes induce both, optical activity and birefringence, at the same time. The permittivity tensor taking into account the birefringence is symmetric and follows the form

$$\boldsymbol{\varepsilon}' = \begin{pmatrix} \varepsilon_{11} + \Delta\varepsilon_{11} & \Delta\varepsilon_{12} \\ \Delta\varepsilon_{12} & \varepsilon_{11} - \Delta\varepsilon_{11} \end{pmatrix}.$$

By rotating the coordinate frame around the *c*-axis, the permittivity can be diagonalized as

$$\boldsymbol{\varepsilon}'' = \begin{pmatrix} \varepsilon_{11} - b & 0 \\ 0 & \varepsilon_{11} + b \end{pmatrix},$$

where $b = \sqrt{\Delta\varepsilon_{11}^2 + \Delta\varepsilon_{12}^2}$.

The optical activity introduces additional imaginary antisymmetric components into the permittivity tensor.

$$\boldsymbol{\varepsilon}''' = \begin{pmatrix} \varepsilon_{11} - b & ic \\ -ic & \varepsilon_{11} + b \end{pmatrix},$$

The coefficient $c$ is the effective optical activity and $b$ is the effective birefringence.

Due to the finite extinction depth $\delta$ of the THz-frequency excitation pulses, the light-induced changes in permittivity decay exponentially, resulting in a depth-dependent non-equilibrium permittivity tensor

$$\boldsymbol{\varepsilon}'''(z) = \begin{pmatrix} \varepsilon_{11} - b(0) \times e^{-\frac{z}{\delta}} & ic(0) \times e^{-\frac{z}{\delta}} \\ -ic(0) \times e^{-\frac{z}{\delta}} & \varepsilon_{11} + b(0) \times e^{-\frac{z}{\delta}} \end{pmatrix},$$

where $b(0)$ and $c(0)$ are the changes in permittivity at the surface $z = 0$.

Since the 5-mm penetration depth of the 808-nm probe pulses is much larger than the sample thickness, the measured polarization rotation signal accumulates when the optical probe propagates through the sample.

We employed the following Jones matrix analysis to differentiate between the signal contributions from the time-dependent optical activity and birefringence(36). At equilibrium, the crystal is isotropic, and the probe pulses remain linearly polarized as they propagate in the sample. For each incident probe polarization, the half waveplate behind the sample is rotated to balance the intensity on the two photodiodes in the absence of the pump.

The excitation pulses change the permittivity as described above. The incident linearly polarized probe pulse can be decomposed into two eigenvectors following the permittivity tensor in the pumped state described above. Defining $f = \frac{c(0)}{b(0)}$, the eigenvectors in the *a-b* plane are two elliptically polarized light waves with opposite ellipticity.

$$\boldsymbol{e_1} = \begin{pmatrix} \frac{i(-1+\sqrt{1+f^2})}{f\sqrt{1+\frac{(-1+\sqrt{1+f^2})^2}{f^2}}} \\ \frac{1}{\sqrt{1+\frac{(-1+\sqrt{1+f^2})^2}{f^2}}} \end{pmatrix} \qquad \boldsymbol{e_2} = \begin{pmatrix} \frac{-i(1+\sqrt{1+f^2})}{f\sqrt{1+\frac{(1+\sqrt{1+f^2})^2}{f^2}}} \\ \frac{1}{\sqrt{1+\frac{(1+\sqrt{1+f^2})^2}{f^2}}} \end{pmatrix}$$

We note that the decay of the strength of the light-induced state over the extinction depth $\delta$ does not change the eigenvectors themselves, but only the difference in the splitting in the refractive indices (see Fig. S4a).

The effective refractive indices for the two eigenvectors are

$$n_1 = \sqrt{\varepsilon_{11} + \sqrt{b(z)^2 + c(z)^2}}$$

$$n_2 = \sqrt{\varepsilon_{11} - \sqrt{b(z)^2 + c(z)^2}}.$$

The phase delay between the two eigenvectors increases with propagation length, with a rate proportional to the difference in refractive index.

$$\frac{d\gamma}{dz} = \frac{2\pi}{\lambda_0}(n_1 - n_2)$$

Since the light-induced change in the effective refractive indices is small compared to the static refractive index ($\sqrt{b^2 + c^2} \ll n_0$),

$$\frac{d\gamma}{dz} \approx \frac{1}{\lambda_0} 2\pi \left( \left( n_0 + \frac{\sqrt{b^2 + c^2}}{2n_0} \right) - \left( n_0 - \frac{\sqrt{b^2 + c^2}}{2n_0} \right) \right) = \frac{1}{\lambda_0} 2\pi \frac{\sqrt{b^2 + c^2}}{n_0}.$$

Integrating over the entire sample thickness $d$, we acquire the phase difference between the two eigenvectors (see also Fig. S4b)

$$\gamma = \int_0^d \frac{1}{\lambda_0} 2\pi \frac{\sqrt{b^2 + c^2}}{2n_0} dx = \frac{\delta}{\lambda_0} 2\pi \frac{\sqrt{b(0)^2 + c(0)^2}}{n_0} \left(1 - e^{-\frac{d}{\delta}}\right).$$

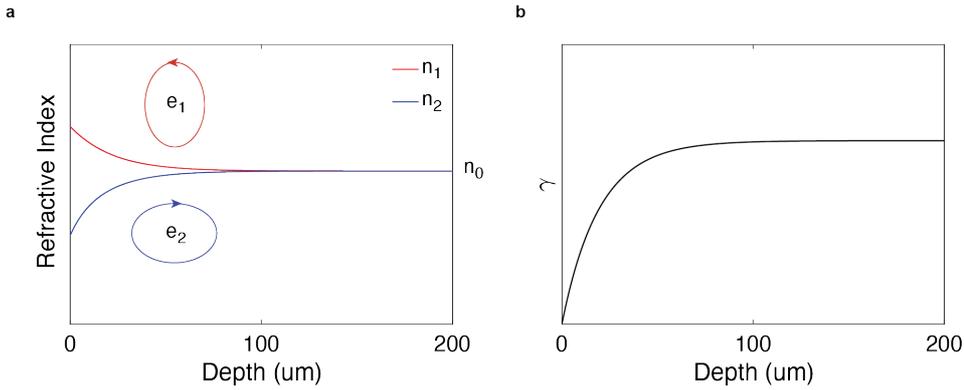

**Figure S4. | Schematic of the light-induced polarization rotation signal. (a) The refractive indices for the two elliptically polarized eigenvectors as a function of propagation depth in the BPO$_4$ sample. The difference between the refractive indices decreases with depth. (b) The accumulated phase difference as a function of the propagation depth.**

The two eigenvectors are then recombined after exiting the sample. The beam passes through the half waveplate and is split onto the two detectors, intensity balanced in the absence of the pump. The polarization rotation signal is proportional to the normalized difference between the intensities on the two detectors

$$\theta(\varphi) = \frac{I_1 - I_2}{2(I_1 + I_2)} = \frac{1}{2\sqrt{1 + 1/f^2}} \sin(\gamma) + \frac{(\cos(\gamma) - 1)}{4(1 + f^2)} \sin(4\varphi - \phi).$$

Here, $\varphi$ is the incoming polarization angle relative to the pump and $\phi$ is the relative angle between the pump to the optical axis of the transient birefringent state. For phase differences $\gamma \ll 1$, we approximate $\sin(\gamma) \approx \gamma$ and $\cos(\gamma) \approx 1 - \frac{\gamma^2}{2}$ and find

$$\theta(\varphi) = \frac{\pi\delta\left(1 - e^{-\frac{d}{\delta}}\right)}{\lambda_0 n_0} c(0) - \frac{\pi^2\delta^2\left(1 - e^{-\frac{d}{\delta}}\right)^2}{2\lambda_0^2 n_0^2} \sin(4\varphi) b(0)^2 .$$

It is clearly seen here that the signal is composed of two contributions. The first one is proportional to the optical activity coefficient $c(0)$ and does not depend on the incoming probe polarization. The second contribution is proportional to the square of the birefringence coefficient $b(0)^2$ and depends on the incoming probe polarization with a 90°-periodicity, as observed in the experiments. The average of the birefringence contribution over all the incident probe polarizations is zero.

Hence, the pure optical activity $c$ is proportional to the average value of the polarization rotation signal over all the incoming polarization of the probe pulse. The pump-induced polarization rotation $\theta(\varphi)$ is fitted with the above expression at each time delay. The fitting result of the experimental data shown in Fig. 3b of the main text is plotted in Figure S5.

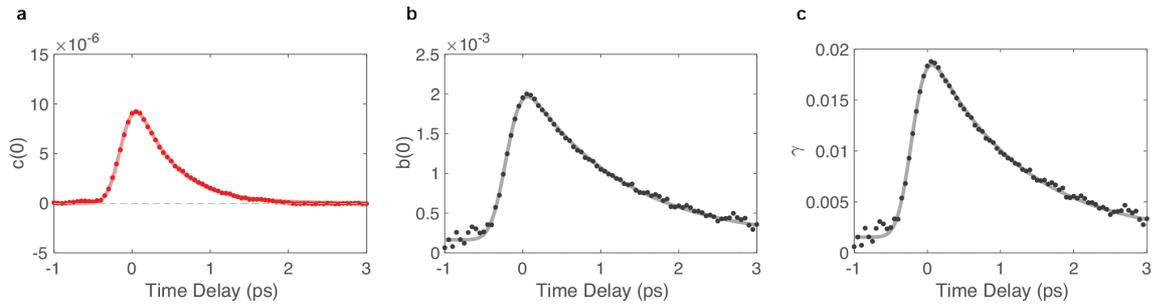

**Figure S5. | Parameter fitting result of data in Figure 3b. (a) Fitting result of optical activity. (b) Fitting result of birefringence. (c) Fitting result of phase difference.**

The rotary power $\rho$ is defined as being proportional to the optical activity coefficient $c$, with

$$\rho = \frac{\pi c}{\lambda_0 n_0} .$$

The birefringence-induced difference in refractive index shown in Figure 3b & 3e of the main text follows

$$\Delta n = \frac{2\pi b}{\lambda_0 n_0}.$$

The light-induced time-dependent rotary power, extracted from our experiments as described above and shown in Figures 3c and 3e of the main text, was then fitted by the functions of the following form, which is a product of a step function with finite rise time $\sigma$ and an exponential decay with decay time $\tau$, reflecting the excitation and decay of the optical phonons by the THz-frequency excitation pulse.

$$\theta(t) = A\left(\mathrm{erf}\left(\frac{t-t_0}{\sigma}\right)+1\right) * e^{-\frac{t}{\tau}}.$$

**S4. Symmetry Analysis of the Nonlinear Phonon Coupling**

Following Neumann's principle, the nonlinear coupling between phonons must be totally symmetric under all operation elements of the point group crystal class(*10*). For BPO$_4$ with point group symmetry $\bar{4}$, the lowest order coupling to the B-symmetry phonon modes is given by

$$U = \alpha\left(Q_{Ey}^2 - Q_{Ex}^2\right)Q_B + \beta Q_{Ex} Q_{Ey} Q_B.$$

In the experiment, we selectively excited only one of the two doubly degenerate E modes at a time, thus the last coupling term with the coefficient $\beta$ is not considered.

We note that in the $\bar{4}$ point group system, E-symmetry modes also drive the system into a transient chiral state (Table S4). However, due to the two-fold rotation symmetry, all the odd-order coupling terms between E-symmetry modes (such as $Q_{Ex}^3$ or $Q_{Ex}^2 Q_{Ey}$) are not allowed, prohibiting any rectified displacement of the crystal lattice along the E-symmetry modes. Therefore, the transient chirality can only be induced by the rectified displacement in the B-symmetry modes.

## S5. Density Functional Theory Calculations

We performed first-principles calculations in the framework of density functional theory (DFT) to explore the phonon excitation spectrum, the anharmonic lattice coupling coefficients, and the optical response of $BPO_4$ within the following technical and numerical settings.

In general, we used the Vienna Ab-initio Simulation Package (VASP) 6.2 DFT implementation(*37-39*) and the Phonopy software package(*40*) for the phonon calculations. Our computations utilized pseudopotentials generated within the Projector Augmented Wave (PAW) method(*41*). Specifically, we took the configured default pseudopotentials for B $2s^2 2p^1$, P $3s^2 3p^3$, and O $2s^2 2p^4$ potentials and applied the Local Density Approximation (LDA) for the exchange-correlation potential. As the final numerical setting, after convergence checks, we used a 7x7x5 Monkhorst-Pack(*42*) generated k-point mesh sampling of the Brillouin zone and a plane-wave energy cutoff of 600 eV for the calculations of structural relaxation and phonons. We re-iterated self-consistent calculations until the change in total energy converged up to $10^{-8}$ eV. All the phonon mode and anharmonic coupling constant computations were performed in a 2x2x2 conventional cell of $BPO_4$. All the calculations of the mode effective charge and the dielectric response, including the optical activity, were performed within the LOPTICS settings of VASP, in addition to the approaches utilized in References(*43, 44*). In order to achieve convergence for the optical properties, we had to increase the k-point mesh to 17x17x11 points and increased the number of empty states to 400.

The starting point of our first-principles investigation is the tetragonal cell of $BPO_4$. First, we determined the DFT equilibrium volume for this cell, which is $V_{cell}$ = 122.8 Å$^3$ (*a* = *b* = 4.30 Å, *c* = 6.62 Å). We found the atomic equilibrium positions at the following Wyckoff positions B 2d, P 2a and O 8g (0.259,0.141,0.127). We then calculated the relevant phonon eigenfrequencies and eigenvectors at the Brillouin zone center, and their respective coupling constants defined in Supplementary Section S5. We list the relevant data in Table S2.

| Phonon symmetry | $\omega_{TO}$ (THz) | $\alpha$ (meV/u$^{3/2}$ Å$^3$) | $Z^*$ (q$_e$/u$^{1/2}$) |
|---|---|---|---|
| E(2) | 18.1 | – | 0.76 |
| B(1) | 15.8 | 1.16 | 0.85 |
| B(2) | 17.6 | 39.93 | 0.37 |
| B(3) | 28.1 | −16.62 | 0.87 |
| B(4) | 31.5 | 105.59 | 1.57 |

Table S2. | Relevant parameters of the optical phonons involved in the light-induced chiral state, obtained from DFT calculations: transverse optical frequency, nonlinear coupling coefficient, mode effective charge.

Lastly, we computed the changes in the symmetric and antisymmetric parts of the dielectric tensor arising from the lattice distortions along the coordinates of the four transiently displaced B-symmetry modes. We list the changes in the optical permittivity $\varepsilon$ arising from the associated birefringence and optical activity in Table S3.

| Phonon symmetry | $\partial\varepsilon_{11}/\partial Q_{B,i}$ (10$^3$ u$^{-1/2}$Å$^{-1}$) | $\partial\varepsilon_{12}/\partial Q_{B,i}$ (10$^3$ u$^{-1/2}$Å$^{-1}$) | $\partial\rho/\partial Q_{B,i}$ (° mm$^{-1}$ u$^{-1/2}$ Å$^{-1}$) |
|---|---|---|---|
| B(1) | 31 | 31 | 15 |
| B(2) | 33 | -8 | -9 |
| B(3) | 5 | 5 | -7 |
| B(4) | -36 | -44 | -43 |

Table S3. | Phonon amplitude dependent changes in the dielectric tensor elements and the optical activity relevant for the light-induced chiral state, obtained from DFT calculations.

## S6. Simulation of the nonlinear phonon dynamics and the transient optical properties

We used the equations of motions, introduced as Eqs. (1) and (2) in the main text, to calculate the crystal lattice dynamics induced by the resonant optical excitation of the doubly-degenerate mode E(2) at about 19 THz frequency. The starting point for all the phonon parameters of this procedure were the first-principles calculations described in Supplementary Section S5.

The left panels of Figure S6 shows these dynamics for the excitation with the optical pump polarized along the crystal $a$ axis. Figure S6a) plots the THz electric field with center frequency of 19 THz and peak field of 5 MV/cm. Figure S6b) plots the oscillations of the resonantly driven E-symmetry phonon $Q_{E,a}$ polarized along $a$. Figure S6c) depicts the response of all the four nonlinearly coupled B-symmetry modes that are rectified due to the force, which is quadratic in the amplitude of $Q_{E,a}$ (see Eq.(2) of the main text). Figure S6d) and e) show the changes of the birefringence and the optical activity, respectively, arising from the combined rectified motion of all the B-symmetry modes and using the data presented in Table S3. Figure S6f) to j) show the equivalent data set for the optical excitation along the crystal *b*-axis. These data sets allowed us to simulate the time-delay dependent changes in the optical properties shown in Figure 3 of the main text.

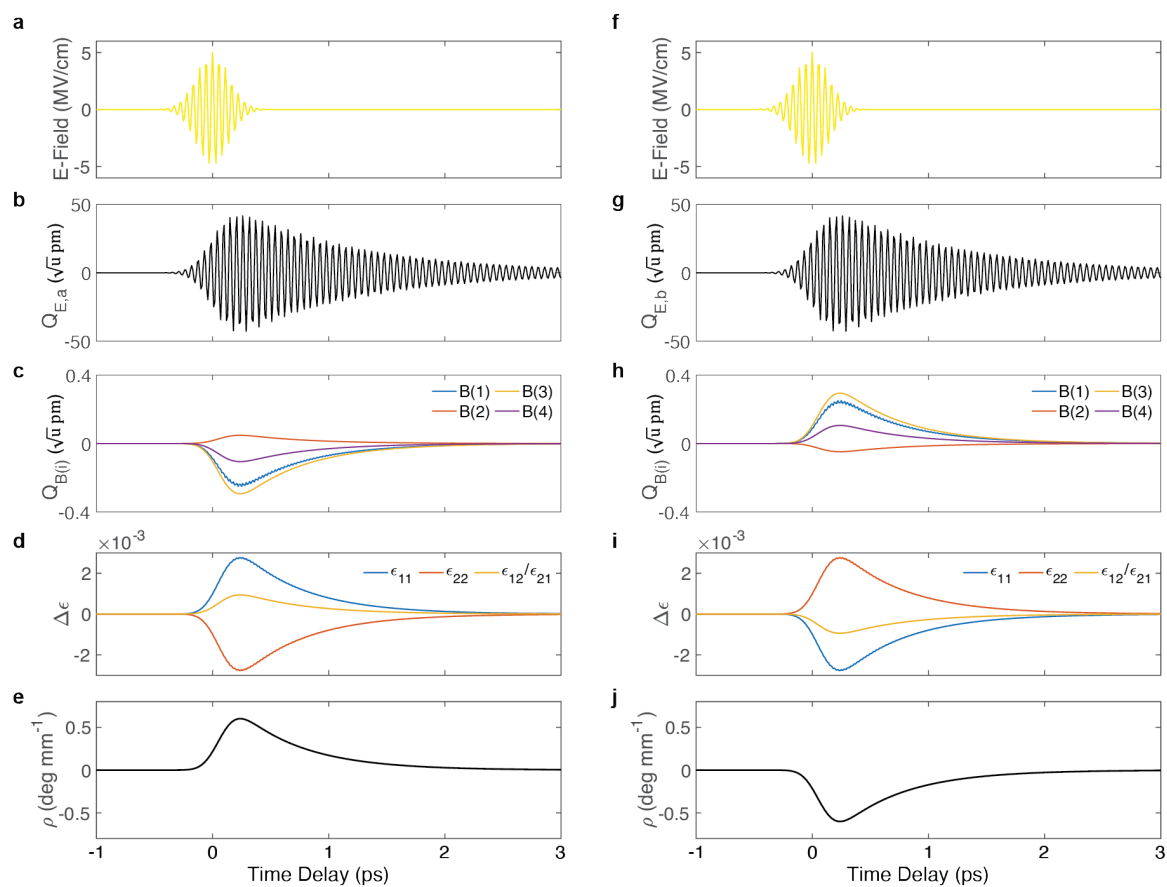

**Figure S6. | Simulation of nonlinear phonon dynamics.** (a) THz Electric field pump. (b) Dynamics of E mode. (c) Dynamics of four B modes. (d) Change in permittivity. (e) Change in rotary power. (f)-(j) Equivalent data set for pumping along the crystal *b*-axis.

## S7. Definition of Antiferro-chiral Systems

The term 'antiferro-chiral' has been previously used in the literature to describe racemic crystals, comprising equivalent ratios of chiral fragments, for example molecules with opposite handedness(*5*). Here, based on symmetry arguments, we provide a general and rigorous definition applicable to both molecular and crystalline systems.

Antiferro-chiral crystalline systems are defined as those achiral systems, which possess at least one vibrational mode that induces chirality in the system. These systems have the potential to form a chiral lattice structure under external stimuli that couple linearly to such vibrational modes.

In crystalline (solid state) systems, materials of different symmetries can be categorized into the 32 crystallographic point groups. For each of the 21 achiral point groups, there exists at least one irreducible representation (irrep) that induces chirality in the system (see Table S4). However, the presence of a phonon mode with such symmetry depends on the specific crystal structure. Hence, only those achiral systems, which host at least one such mode, are antiferro-chiral. The relationship of the classification is shown in Figure S7. It is important to note the following properties of antiferro-chiral.

- Antiferro-chirality is not an intrinsic property of a point group or a space group.
- Antiferro-chirality exists in both, centrosymmetric and non-centrosymmetric systems.

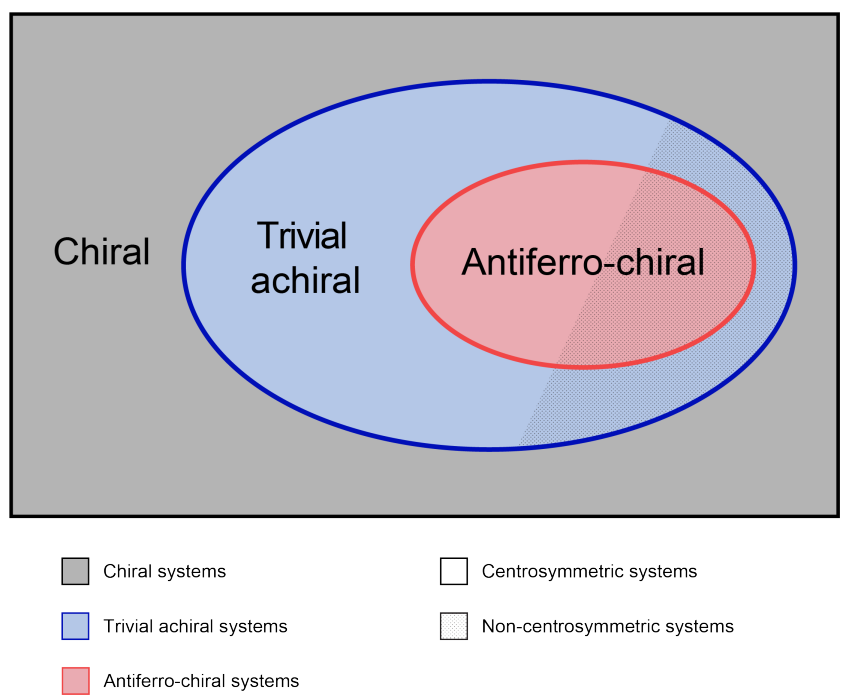

Figure S7. | Classification of material systems based on symmetry. Antiferro-chiral is a subset of achiral systems.

| Non-centrosymmetric Achiral Point group | Irreps | Centrosymmetric Achiral Point group | Irreps |
|---|---|---|---|
| m | A'' | -1 | $A_u$ |
| mm2 | $A_2$ | -3 | $A_u, E_u$ |
| 4mm | $A_2, E$ | 2/m | $A_u$ |
| 3m | $A_2, E$ | 4/m | $A_u$ |
| 6mm | $A_2, E_1, E_2$ | 6/m | $A_u, E_{2u}$ |
| -42m | $A_2, B_1, E$ | mmm | $A_u$ |
| -6m2 | $A_1'', E''$ | 4/mmm | $A_{1u}$ |
| -43m | $A_2, E, T_1, T_2$ | -3m | $A_{1u}, E_u$ |
| -4 | B, E | 6/mmm | $A_{1u}, E_{2u}$ |
| -6 | A'', E'' | m-3 | $A_u, E_u, T_u$ |
|  |  | m-3m | $A_{1u}, E_u, T_{1u}, T_{2u}$ |

**Table S4. | Irreducible representation that induces chirality in the 21 achiral crystalline point groups.**